# ON THE USE OF SMART ANTS FOR EFFICIENT ROUTING IN WIRELESS MESH NETWORKS

Fawaz Bokhari and Gergely Zaruba

Department of Computer Science and Engineering, The University of Texas at Arlington, Texas, US
fawaz.bokhari@mavs.uta.edu, zaruba@cse.uta.edu

## ABSTRACT

*Routing in wireless mesh networks (WMNs) has been an active area of research for the last several years. In this paper, we address the problem of packet routing for efficient data forwarding in wireless mesh networks (WMNs) with the help of smart ants acting as intelligent agents. The aim of this paper is to study the use of such biologically inspired agents to effectively route the packets in WMNs. In particular, we propose AntMesh, a distributed interference-aware data forwarding algorithm which enables the use of smart ants to probabilistically and concurrently perform the routing and data forwarding in order to stochastically solve a dynamic network routing problem. AntMesh belongs to the class of routing algorithms inspired by the behaviour of real ants which are known to find a shortest path between their nest and a food source. In addition, AntMesh has the capability to effectively utilize the space/channel diversity typically common in multi radio WMNs and to discover high throughput paths with less inter-flow and intra-flow interference while conventional wireless network routing protocols fail to do so. We implement our smart ant-based routing algorithm in ns-2 and carry out extensive evaluation. We demonstrate the stability of AntMesh in terms of how quickly it adapts itself to the changing dynamics or load on the network. We tune the parameters of AntMesh algorithm to study the effect on its performance in terms of the routing load and end-to-end delay and have tested its performance under various network scenarios particularly fixed nodes mesh networks and also on mobile WMN scenarios. The results obtained show AntMesh's advantages that make it a valuable candidate to operate in mesh networks.*

## 1. INTRODUCTION

Wireless mesh networks (WMNs) have emerged as a promising technology for the provision of last mile broadband Internet access infrastructure in both urban and rural environments. Such networks are characterized as fixed backbone WMNs where relay nodes are generally static and are mostly supplied by permanent power sources [1]. With the availability of off-the-shelf, low cost, commodity networking hardware, it is possible to incorporate multiple radio interfaces operating in different radio channels on a single mesh router; thus forming a multi-radio mesh network. This enables a potential large improvement in the capacity of mesh networks compared to single-radio mesh networks [20]. However, due to the shared nature of the wireless medium, nodes compete with each other to access the wireless channel, resulting in possible interference among nodes which may affect traffic in the network. Furthermore, the fact that nodal density in a typical WMN is high compared to other wireless networks, make the flow interference issue in WMNs more severe. Incorporating techniques to specifically address these characteristics in a WMN routing protocol could improve the overall network flow capacity or performance of individual flows in the network.

Generally, routing algorithms can be defined as multi-objective optimization problems in a dynamic stochastic environment. However, formalizing routing as such optimization problem requires complete knowledge of traffic flows between each node in the network; this is prohibitively difficult to model in the presence of rapidly changing network dynamics (found in typical WMNs). Therefore, heuristic policies are often used to create quasi-optimal routing in WMNs. There has been a significant body of research in designing efficient heuristic based routing protocols and metrics for WMNs (for a quick overview see [2 - 5]). In the wired networking domains, a new family of routing algorithms has been proposed based on swarm intelligence by Dorigo et al. called Ant Colony Optimization (ACO) framework [6, 7], which is a meta-heuristic approach for solving hard optimization problems.

ACO algorithms draw their inspiration from the behaviour of real ants, which are known to find the shortest path between their nest and a food source by a process where they deposit pheromones along trails (acting like a local message exchange in a communication network). Ants generally start out moving at random, however, when they encounter a previously laid trail, they can decide to follow it, thus reinforcing the trail with their own pheromone substance. This process is thus characterized as a positive feedback loop, where the probability with which an ant chooses a path increases with the number of ants that previously chose the same path [7]. Hence, In ACO framework, *artificial ants* probabilistically and iteratively converge to a solution by taking into account pheromone trails deposited by other ants and other local heuristics. Artificial ants move stochastically (instead of deterministically) in the solution space, therefore they can explore a wider variety of possible solutions of a problem independently and in parallel. A more detailed explanation of the ACO framework can be found in [6].

## 1.1 Problem Addressed

In this paper, we address the problem of packet routing for efficient data forwarding in wireless mesh networks. Since most of the traffic in a mesh network usually flows between regular nodes and a few Internet gateways (i.e., rarely end-to-end between regular nodes). This can result in an uneven loading of links and can cause certain paths to be saturated. Similarly, the existence of inter-flow interference among the nodes and intra-flow interference within a transmission path may affect traffic loads on mesh nodes in a multi-radio WMN. The main objective of any mesh routing protocol is thus to effectively distribute the traffic by selecting channel diverse paths with less inter/intra flow interference.

## 1.2 Our Contributions

The salient features of our work that set us apart from the existing routing protocols in WMNs are listed as follows:

- We propose a distributed routing mechanism which enables the use of smart ants to probabilistically and concurrently perform the routing and data forwarding in order to stochastically solve a dynamic network routing problem.
- We formally define the properties and conditions necessary in the design of such smart ant based routing algorithm for WMNs. These smart ants help to effectively utilize the space/channel diversity typically common in multi radio WMNs.
- One interesting result is that smart ants has the capability to discover high throughput paths with less inter and intra-flow interference when conventional wireless network routing protocols fail to do so.
- The stability of any routing protocol depends upon how quickly it adapts itself to the changing dynamics of the network. We demonstrate through simulations that AntMesh quickly converges to the best path under situations when traffic characteristics change. We tune the parameters of AntMesh algorithm to study the effect of these on the performance in terms of the routing load and end-to-end delay.
- To the best of our knowledge, our proposed forwarding technique is among the first works that investigates the use of smart ants in WMNs and demonstrates a possible good performance advantage.

### 1.3 Paper Organization

The rest of the paper is organized as follows. We start with providing an overview of some of the existing ACO based routing protocols in wireless networks in Section II. The concept of smart ants in mesh networks and the necessary properties that they should possess are discussed in Section III. We will explain the working of our smart ant-based routing algorithm (AntMesh) in Section IV together with the description of how smart ants are used in AntMesh to capture the traffic load and inter/intra flow interference in WMNs. We describe the implementation details of AntMesh followed by simulation results to evaluate its performance to existing routing schemes in Section V. Finally we conclude the paper in Section VI.

## 2. RELATED WORK

There have been research efforts in using ant-based techniques for efficient data forwarding in wired networks [8, 9, 13] and mobile ad-hoc networks (MANETs) [10-12, 25-28, 34, 35]. This section provides an overview of some of the more prevalent ant-inspired routing algorithms in communication networks.

*AntNet:* one of the first applications of ant colony optimization framework (ACO) for wired network routing is AntNet [8]. In AntNet, routing is achieved by generating forward ants at regular intervals from a source node to a destination node to discover a low cost path and by backward ants that travel from destination to source to update pheromone table at each intermediate node. Forward ants keep track of trip times from source to destination node using the data traffic queues in order to experience the same delay that data packets experience. Forward ants select the next hop by a probabilistic decision rule which takes into consideration the pheromone intensity which is reinforced by other backward ants and heuristic information which is based on the queue length of the intermediate node. Once, a forward ant reaches the destination node, a backward ant is generated that tracks back to the source using high priority queues (for timely delivery) reinforcing the selection probability of intermediate nodes according to the fitness of the trip times of forward ants. However, the fact that AntNet was proposed for wired networks makes it unsuitable for WMNs because of their unique characteristics i.e. wireless interference and load balancing etc. etc. There are variations and extensions, e.g., [13, 14], of the original AntNet algorithm which targeted wired networks and thus are outside the scope of our research.

*AntHocNet [12]*: is a hybrid, multi-path, ant based algorithm. It consists of both a reactive and a proactive component. The reactive part is used for route establishment whereas the proactive component is used for route maintenance. The reactive component is used at the start of a data session where a *reactive forward ant* is broadcast to find multiple paths to the destination and upon reaching the destination; a *reactive backward ant* sets up the multiple paths to the destination using local heuristic information. While data packets are being routed, proactive forward ants are also generated. This helps in exploring new paths and getting up-to-date link quality information. One of the drawbacks of AntHocNet is the number of ants that need to be sent over the network for establishing routes to destinations as they are broadcast during a route discovery phase. Also, each ant stores list of visited nodes from source to current node and depending upon the distance to the destination, this list (and thus the ant's size) can grow long, increasing routing overhead.

*POSANT*: the authors of [10] present a position based ant colony routing algorithm for MANETs, in which they make use of position based instruments (e.g., GPS receivers) to combine with the ACO technique. POSANT is a reactive routing algorithm and [10] argues that the use of position information can greatly reduce the number of ant generations while reducing the route establishment time as well. Position information is used in the heuristic maintained at each node helping ants decide what next hop to take in the path discovery phase. However,

POSANT makes an assumption that the transmission time of an ant to its neighbours is the same for all nodes therefore ignoring packet loss and flow interference, which are quite common in wireless networks.

*SARA*: Rosati *et al.*, [34] have proposed a distributed ant routing algorithm for ad hoc networks where nodes are frequently joining and leaving the network and minimum signalling overhead is required. One of the design objectives of their proposed approach was to have low computational complexity and for this particular purpose, the algorithm creates routes on-demand making it a reactive protocol. Moreover, ants in SARA algorithm store only the node identity information to avoid extra control overhead and only the pheromone value is taken to make a forwarding decision. The authors have conducted extensive simulations and showed the effectiveness of their algorithm in ad hoc networks with critical connectivity when compared with AODV. Since the approach aims at minimizing complexity in the nodes making it as simple as possible, the authors claim that SARA can be helpful in achieving seamless routing in heterogeneous networks. However, the same simplicity comes with the drawback that it does not capture the network dynamics in terms of link characteristics and interference and taking necessary steps accordingly. Therefore it is not suitable for networks with high nodal densities and traffic flows moving to mostly one side of the network (gateways) typically like in wireless mesh networks.

*DAR*: recently, the authors of [35] have presented a simple ant routing algorithm for ad hoc networks with the objective of minimizing the protocol overhead. To achieve this objective, they have proposed a control neighbour broadcast method in which, a broadcast ant is unicast to the second-hop neighbour. Similarly, data packets are used to refresh paths of the active sessions in order to reduce the control overhead. Their third approach is implemented in link failure situation to repair that link locally using the two ends of the link rather than initiating a new source to destination path request. They have conducted extensive simulations to prove the main objective of their proposed approach i.e. less signalling overhead when compared with other existing ant-based routing algorithms.

For wireless mesh networks, we presented an interference-aware routing scheme based on ACO framework for single and multi radio WMNs [15, 29]. With respect to the previously quoted works on AntMesh, this paper presents a deeper comparison analysis and insight on the performance of smart ants in WMNs. We study in depth the working of AntMesh routing algorithm and describe the desirable properties of smart ants particularly suited for multi radio WMNs. We demonstrate that the network performance can be optimized when AntMesh algorithm parameters are carefully selected. The use of smart ants for routing on mesh networks shows a noticeable advantage against some well-known existing biological routing approaches.

## 3. SMART ANTS IN MESH NETWORKS

This section explains the rational on the use of ants for routing in WMNs. We will explain the concept of the smart ants, how they differ from regular ants used in previous techniques and the necessary properties that these smart ants should possess in order to efficiently perform routing in WMNs.

### 3.1 Desirable Properties of Smart Ants

Let us consider what kind of ants would be suitable for creating network paths in defining an ant based routing algorithm for MWNs. We argue that a routing algorithm based on smart ants designed for mesh networks should have the following desirable properties:

- The smart ants while creating paths should take into account the two types of interferences that inherently exist in mesh networks namely inter-flow interference among the nodes and intra-flow interference along the path of a flow.

- The smart ants should be able to evaluate the load on nodes in order to properly qualify the outgoing links. This would help in *detouring* the packets to new route and hence would result in a more load balanced network.
- Since network nodes in a WMN can be equipped with multiple radios, a smart ant should be able to discover *channel-diverse paths* in order to reduce the interference and effectively improve the overall network throughput.

Smart ants are designed to exhibit the above mentioned desirable properties and it is because of these properties which make our ants smart (intelligent) and that is why we call them *smart ants*. In the following, we show how to incorporate these desirable properties in our smart ant-based routing algorithm (AntMesh).

## 4. THE ANTMESH ALGORITHM

In this section, we describe the details of AntMesh; a distributed routing algorithm which incorporates smart ants to find high throughput paths with less interference and improved load balancing specifically designed for WMNs. The basic operation of AntMesh follows the routing protocol described in [15, 29]. Smart ants in the form of control packets are generated at regular intervals from each node towards destinations in the network. Indeed, three types of ants are generated: forward smart ants (FSA) which travel from source to destination to discover paths, backward smart ants (BSA) travelling from the destination to the source to update the routing tables and hello smart ants (HSA) which collect the local link quality information to populate link estimation table. In AntMesh, both the FSA and BSA use high priority queues so that the FSA do not need to carry their per hop experienced trip times, rather BSA will estimate their trip time.

### 4.1 Data Structures

In AntMesh, every node maintains three types of data structures explained as below.

*Pheromone table (Probabilistic routing table).* This data structure stores the fitness of choosing a specific neighbour as next hop to reach a particular destination in the form of a probability. In other words, it contains pheromone trail information for routing from current node to destination node via next hop. Thus, the pheromone table at a particular node $k$ contains $m_k$ rows where $m_k = |N_k|$ ($N_k$ is the set of neighbouring nodes to node $k$) and each row contains $N$ columns, where $N$ is the total number of possible destinations in the network (total number of nodes, or population). So an entry $P_{id}$ is the probability of sending a packet to destination $d$ via link $i$ and thus following relation holds for every column in the pheromone table at node $k$:

$$\sum_{i \epsilon N_k} P_{id} = 1 \qquad \forall\, d\, \epsilon\, [1 \ldots N] \tag{1}$$

*Delay table.* The second data structure that is maintained by each node is the delay table that stores the average trip time to each destination in the network from the current node (thus this table will have $m$ entries one for each possible destinations in the network). The value stored is an average calculated from the delay value carried by the last $W$ number of smart ants received.

*Link estimation table.* The third table maintained by AntMesh is the local estimation table which contains the quality/strength of the outgoing links of that particular node to its neighbours. AntMesh uses hello smart ants (HSA) to measure these local link statistics in terms of link level packet transmission delays.

A more detailed explanation on how these data structures are populated will be provided later in this section.

### 4.2 Node Transition Rule

In order to increase the chances of selecting paths with less interference and more throughput, AntMesh has adopted a pseudo-random node transition rule. A forward smart ant $v$ at an intermediate node *k* chooses the next hop *u* to reach to a particular destination *d* according to Eq. (2) and Eq. (3).

$$u = \begin{cases} \arg\max_{u \in N_k}\{\tau_v(d,u)\} & if\ p \leq p_0 \\ P_v(d,u) & otherwise \end{cases} \quad (2)$$

$$P_v(d,u) = \frac{\tau_v(d,u)}{\sum_{i \in N_k} \tau_v(d,i)} \quad (3)$$

where *p* is a random number uniformly distributed between [0, 1] and $p_0$ is a constant in the range [0, 1]. Similarly $\tau_v[d,u]$ is the pheromone intensity on the link connecting next hop *u* with *k* to reach the destination *d*. Eq. (2) indicates that if $p \leq p_0$, the node with the maximum pheromone value among the neighbours will be selected, otherwise, a proportional selection will be made with the probability $P_v(d,u)$ based on the probability distribution of the neighbours as shown in Eq. (3). Note that the parameter $p_0$ determines the relative importance of exploitation versus exploration. In AntMesh, since pheromone table is the only data structure that contains the routing information, all the data packets are forwarded based on this pseudo-random node transition rule. Therefore, the high value of $p_0$ would direct all the traffic to the best path as the link with maximum pheromone value would be selected most of the time and a lower value of $p_0$ will allow the data traffic to be spread across multiple links thereby resulting in automatic load balancing. How much should this $p_0$ be set depends highly on the offered load on the mesh network and we will study the effect of this parameter to AntMesh performance in detail in our simulation section. Setting $p_0$ values to 1 will always forward the packets on links with the best quality.

### 4.3 Interference Estimation Rule

This rule is designed to fully capture the characteristics of WMNs including the types of interferences that exist in such networks and to incorporate the desired properties of smart ants in AntMesh algorithm. It consists of two modules i.e. link estimation module (LEM) and path estimation module (PEM). These two modules help the smart ants to accurately measure the inter and intra-flow interference in the network.

### 4.3.1 Link Estimation Module (LEM)

The link estimation module of AntMesh calculates the quality/cost of a wireless link in terms of average transmission time (delay) it takes the MAC layer to send a packet on a particular outgoing link. We define this *link transmission delay* as the time from when a packet starts to be serviced by the MAC layer to the instant that it is successfully transmitted (thus including the time required by any or all retransmissions). Let $T_i$ denote the transmission delay over link *i* and $N_{tx}$ denote the number of transmissions including retransmissions needed to successfully receive a packet. (In most practical cases, $N_{tx} \cong 1$, as almost all the current wireless devices are equipped with multi-rate feature which automatically adjusts the link rate according to the link quality, resulting in successful transmission of around 90% packets at the first time - see [18]). $L_{pkt}$ is the data packet size and $R_s$ is the link speed. The link transmission delay can be defined as follows:

$$E[T_i] = N_{tx} \times \left(MAC_{oh} + \frac{L_{pkt}}{R_s}\right) \quad (4)$$

where $MAC_{oh}$ is the standard packet sequence of sending a data packet; in IEEE 802.11, $MAC_{oh}$ is calculated as follows:

$$MAC_{oh} = T_{rts} + T_{cts} + 3T_{sifs} + T_{difs} + T_{ack} \quad (5)$$

and $T_{rts}$, $T_{cts}$, $T_{ack}$ are the times required for the transmissions of RTS, CTS, and ACK frames respectively, $T_{sifs}$ and $T_{difs}$ are the inter-frame spaces: SIFS and DIFS.

However, this delay $E[T_i]$ in Eq. (4) is just the MAC layer transmission delay which does not take the traffic load into account, i.e., the queuing delay, which depends on the number of packets waiting in the buffer for transmission. The packets that are already in the queue must be served by the MAC layer before the new packet that has just arrived at the node would be served. Therefore, the link estimation module (LEM) calculates the total transmission delay of a packet on a particular outgoing link as the packet transmission delay of the number of packets in the buffer plus the transmission delay of the newly arrived packet. This can be shown as below:

$$LQ_i = E[T_i] \times Q_k + E[T_i] \qquad (6)$$

Where $Q_k$ denotes the queue size of node $k$ and thus $LQ_i$ is the link quality of link $i$. Note that $LQ_i$ includes the load on node $k$ and is therefore the total time it would take for a newly arrived packet at a node $k$ to be transmitted to the next hop.

### 4.3.2 Path Estimation Module (PEM)

The path estimation module of AntMesh has been designed to meet the remaining desirable properties of smart ants i.e. discovering channel diverse paths and inter/intra-flow network interference.

After the completion of path discovery of forward smart ant (FSA), i.e. it has reached the destination, a backward smart ant (BSA) is created and is sent back to the source to update the data structures maintained by each node. During the backward smart ant's (BSA) travel to the source, each node that encounters the BSA updates its delay table data structure. They do so by estimating the smart ant's trip time from the current node to the destination node. This trip time is estimated as the sum of the average transmission time it takes to send a data packet from the current node to the next hop node (from where the smart ant has arrived) and the estimated accumulated trip time calculated so far by the BSA. The local link transmission delay of the next hop node is provided by the link estimation module from the link estimation table.

*Inter-flow Interference.* Since, nodes transmitting on the same wireless channel compete for the shared medium, whenever a node is involved in a transmission; its neighbouring nodes should not communicate at the same time with other nodes on the same channel. The PEM of AntMesh captures this type of interference (inter-flow interference) on each node and incorporates it into the local link estimation table. Let $I_{(k)}$ denote the set of queue sizes of node $k's$ interfering nodes. We define $IFLD_{k,i,c}$, "inter-flow link delay" as follows:

$$IFLD_{k,i,c} = LQ_i \times max[I_{(k)}] \qquad (7)$$

where $IFLD_{k,i,c}$ is the link transmission delay of a particular link $i$ transmitting on channel $c$ when inter-flow interference is considered. The path estimation module (PEM) only takes the maximum queue size of the neighbouring node among all the neighbours (because a node with a very short queue length can still be congested if its interfering nodes have a lot of packets to send out) as it really depends on the current activities of all neighbouring nodes to find out if they are indeed contending or not. AntMesh uses hello smart ants (HSA) to capture these node statistics in terms of queue sizes among the neighbouring nodes. Therefore a high-contention link that would result in increased link transmission delay is detected timely by the PEM module of AntMesh.

The rationale behind using the queue lengths of the neighbouring nodes to measure the contention on a link belonging to a particular node can be understood with the help of a simple example as illustrated in Figure 1: Let us assume that each node is configured to use a single channel. Let us say that $S$ is the source node and $D$ is the destination node. So there are two paths to reach to $D$: *S-A-D* and *S-C-D*. In this particular diagram, node $S$ calculates the inter-flow interference of its neighbouring nodes $A$ and $C$ by taking into account the queue sizes of the neighbouring nodes of $A$ and $C$ on links $i$ and $j$ respectively. It is clear that neighbouring node $A$ has more neighbours than node $C$, which initially gives the impression that the contention on link $i$ is going to be high than link $j$. However, it really depends upon the number of packets waiting to be served in the queues of those neighbouring nodes which describes the contention along the link. So the neighbours around link $j$ which are less as compared to link $i$ have more packets waiting in the queue than the neighbours of link *i*. Therefore, the contention on link *i* is less as compared to link *j* which is captured by our estimation module and this would eventually result in improving the overall network performance of a mesh network.

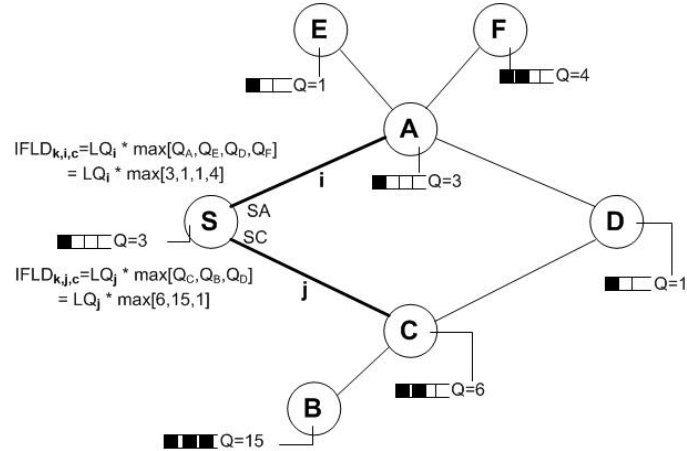

Fig. 1. AntMesh inter-flow interference calculation at node *S*

*Intra-flow Interference.* The backward smart ant (BSA) in our smart ant-based routing follows a deterministic path from destination to source; it keeps a record of the channels of the links on which it has traversed together with the accumulated trip time. When the BSA is received by a node (intermediate, or destination) the PEM checks the smart ant for the links it has traversed; if the two latest hops used the same radio channel then it increases the total transmission time calculated in Eq. (7) by a cost factor $\alpha$ as shown below:

$$ITT_{i,c} = IFLD_{k,i,c} \times \alpha \qquad (8)$$

where $ITT_{i,c}$ (inter + intra-flow transmission time) is the total link transmission delay of a link *i* on channel *c* of a particular node. The cost factor $\alpha$ is calculated as:

$$\alpha = \frac{2Q_{next}L}{B_i} \qquad (9)$$

where $Q_{next}$ is the number of packets waiting in the queue at the next hop node, *L* is the packet size and $B_i$ is the bandwidth of the link. Since, only one link can be active at any time if both the consecutive links are on the same channel then the effective bandwidth will be $B_i / 2$ (i.e., the transmission time should be doubled for packets using the same channel on consecutive links). Similarly, no cost will be added to the total transmission delay when the two last-hop channels are different along the path ($\alpha = 0$). $ITT_{i,c}$ will be added to the trip time that the BSA has accumulated so far from the next hop neighbour node *j* to the destination node *d* denoted as $Trip_{i,d}$. Thus the BSA updates the trip time field of the delay table and hence the total trip time from a node *i* to destination node *d* is cumulative and is calculated as:

$$Trip_{i,d} = ITT_{i,c} + Trip_{j,d} \qquad (10)$$

where $Trip_{j,d}$ is the trip time from next hop node *j* to the destination *d* and is taken from the *Trip Time* field of the BSA packet structure.

### 4.4 Pheromone Updating Rule

After the calculation of trip times $Trip_{i,d}$, BSAs trigger an update of the pheromone table values corresponding to a specific destination. However, there is a need for a mechanism to correctly integrate the transmission time measured by the AntMesh, into the pheromone table entries. This will improve the routing decision process with respect to the quality of the link. Let us

124

International Journal of Wireless & Mobile Networks (IJWMN) Vol. 4, No. 2, April 2012

denote the average trip time for a particular destination *d* stored in the delay table at node *i* with $T_{i,d}$, and the current calculated trip by $Trip_{i,d}$ then:

$$\Delta p = \frac{1}{2} \times \left(\frac{T_{i,d}}{Trip_{i,d}}\right) \quad (11)$$

where $\Delta p$ is the reinforcement value that will be added to the pheromone table and it depends upon how good $Trip_{i,d}$ is. Since, AntMesh keeps up to date local link quality information, the trip time $Trip_{i,d}$ depicts the latest network condition in terms of traffic load and congestion.

Now, the pheromone values in the pheromone table corresponding to a particular destination *d* via next hop *i* can be updated as follows:

$$P_{i,d} = \frac{P_{i,d} + \Delta p}{1 + \Delta p} \quad (12)$$

Similarly, the other neighbors as next hops *j* for the same destination can be downgraded:

$$P_{j,d} = \frac{P_{j,d}}{1 + \Delta p} \text{ where } j \neq i \quad (13)$$

Note that Eq. (12) and Eq. (13) satisfy Eq. (1).

### 4.5 An Illustrative Example

We use a simple wireless network in Figure 2 to illustrate how our proposed smart ant-based routing algorithm works. We denote the source and destination nodes by *S* and *D* respectively in this example. *CH* represents the channel number on which a particular link is configured. The number of packets waiting in the queues are represented next to each node circles. Similarly, the link estimation and delay tables of each node are shown in the figure and the path computation equation is shown above each table to calculate the trip time. Our goal is to find a path from *S* to *D*; the path discovery process works as follows: in order to compute our path metric when the forward smart ant (FSA) traverses the network, we need the link metric for each link traversed and the channel of last two hop links in which they are operating. So we overload the smart ant packet to carry the link metric and the channel of links traversed. Following are the sequence of steps performed by the AntMesh algorithm.

1. Node *S* initiates the route discovery by generating an FSA, which reaches the destination *D* using the pheromone tables on each node by applying pseudo-random node transition rule.
2. The FSA can reach to destination node *D* through two paths i.e. *S-A-F-D* (path 1) or *S-G-C-F-D* (path 2). Now, destination node *D* generates the backward smart ant (BSA) which follows the same path as FSA did and would update the pheromone table, local link estimation table and delay table using our custom designed estimation modules of interference estimation rule.
3. When the BSA reaches node *F*, it records both inter- and intra-flow interferences of the link it has traversed by taking the maximum of the queue sizes of the neighbouring nodes of *D* (max[$Q_D$]=1) and adding $\alpha$ ($\alpha = 0$) if the last two hops link channels are same respectively. It then incorporates these measurements into the link quality of the link (link estimation module - maintained by link quality table). This step will be the same for either path-1 or -2.
4. For path 1 (*S-A-F-D*), as the backward smart ant travels to *A*, it carries with it the accumulated trip time of *F-D* which it calculated on the previous step. On reaching node *A*, the inter-flow interference is measured (max[$Q_F,Q_C,Q_D,Q_E$]=5) and then is incorporated into the local link quality of the link A-*F* stored at link quality table using Eq. (7). Also, since the last two hops' link channels are different (AF = 2, FD = 4), therefore no intra-flow interference exists along these links ($\alpha = 0$). The same calculations are carried for the BSA traversing on path 2 (*S-G-C-F-D)* i.e. on reaching node *C*, the trip time for a packet to reach node *D* from *C* is calculated using (12). This trip time is then integrated into the pheromone table using (14, and 15).

5. Step 4 is repeated for node *G* on path 2 and eventually upon reaching source node *S*, it captures the inter- and intra-flow interferences of its outgoing link on which the smart ant has arrived (updating the data structures accordingly). Note that for path 1, there exist intra-flow interference along the links *SA* and *AF* as they both are on the same channel and this has been taken care by the interference estimation rule using its path estimation module by adding a factor $\alpha$ (fixed to 0.5 in the figure) calculated using Eq. (8).

The example points out that although path-1 contains less number of hops than path-2, the contention in terms of inter- and intra-flow interference along the path is higher. Therefore, the trip time to reach destination *D* on node *S* via the outgoing link *G* is less than through neighbour node *A*. This is depicted in the delay table of node *S* in figure 2. This concludes our example on how our smart ants can capture interferences along the routes; we will show that this behaviour will eventually result in an overall improved performance.

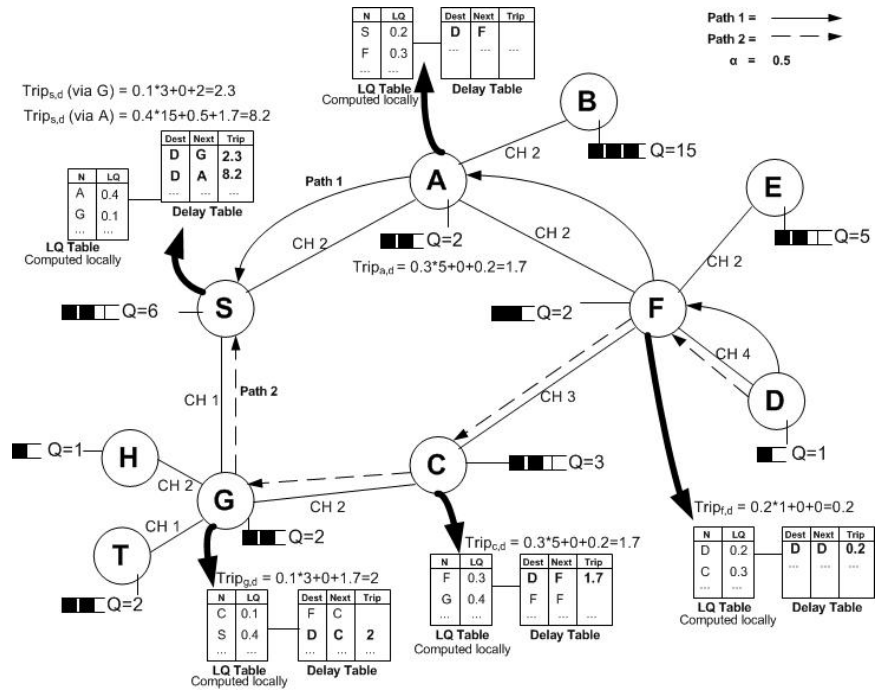

Fig. 2. Illustrative Topology - An example illustrating the working of Smart Ants in AntMesh. The backward smart ant computes trip time and updates delay and pheromone table accordingly. The trip time on each node is calculated using equation: $Trip_{i,d} = LQ_i \times \max[I_k] + \alpha + Trip_{j,d}$ where $\alpha=0.5$ and $Trip_{j,d}$ is the trip time from next hop node to destination, taken from BSA packet

## 5. PERFORMANCE EVALUATION

In this section, we discuss the implementation, setup, motivation behind the setups, and results of our experiments. Our evaluations are divided into three parts. First, we study the stability of our smart ant-based routing algorithm in terms of how quickly it manages to find new paths in dynamic networks by tuning the algorithm parameters in order to reach an optimized behavior. We also discuss the effect of smart ant generation rate on the overall performance of the mesh network by measuring normalized routing load (NRL) and packet delay. The second part of our evaluation consists of comparing AntMesh with some of the existing well-known routing protocols in WMNs using different network topologies and traffic characteristics. Finally, the performance of AntMesh in mobile WMNs where some of the nodes are static and some of them are mobile is compared in the third part of our evaluation.

## 5.1 Implementation

In this subsection we will elaborate the implementation details of our proposed smart ant-based routing algorithm (AntMesh) in ns-2 [19]. We have selected ns-2 for implementation and simulation purposes based on its popularity and acceptance in the academic community. We have implemented our AntMesh routing agent as smart ants which are small fixed size control packets being sent periodically for finding paths and building routing and pheromone tables along the way in the network. In order to measure the local link qualities by our link estimation module (LEM) of AntMesh, a list of one-hop neighboring nodes and their corresponding link qualities (i.e., in the form of link delays) are maintained in a local link estimation table by periodically broadcasting hello smart ants (HSA). Upon receiving the HSA, each node sends back its queue size along with the list of its one-hop neighbors and their corresponding queue sizes to calculate the inter-flow interference by the path estimation module (PEM) of link interference rule. This effectively enables each node to capture 2-hop neighborhood information. Periodical HSA type hello exchange has been widely adopted by most of the existing routing protocols [2, 3, 16, 31].

Since current ns-2 distribution does not support multi-channel communication and wireless nodes having multiple interfaces, we included multiple interface support in ns-2 using [36] document by ramon and campo in order to provide support for multi-radio mesh networks and to evaluate the algorithms on such scenarios. Similarly, we have implemented some of the most recent ant-based routing algorithms i.e. SARA [34] and DAR [35] in ns-2 for comparative study. Our selection of these two protocols for comparative evaluation is because these are the most recent routing algorithms proposed that are based on ants to build routing paths. Although, both SARA and DAR are designed for MANETs but the fact that AntMesh is the first ant-based routing approach to deal with WMNs to the best of our knowledge, force us to choose these two protocols. However, this does not imply that MANETs and WMNs have similar characteristics and therefore, a need to design a new routing algorithm for WMNs specially tailored for its unique characteristics was required as mentioned in section I.

Since, most of the traffic in a real WMN is either to or from a wired network [1, 20] (i.e., through Internet gateway points), in our simulations, flows are destined to one to four gateway nodes. The common configuration parameters for all simulation studies in this section are listed in Table II. Although some may argue that some of these values are low (e.g., link bandwidth), our purpose is to provide a comparative evaluation (relative results) and thus we believe that these values will not strongly influence our results. Each of our depicted data points in the results is an average over enough simulation runs to claim a 95% confidence that the relative error of them is less than 5%. Any topology related change in the simulation will be mentioned in the appropriate subsection.

**TABLE II**
SIMULATION CONFIGURATIONS

| Simulation area | 1000 x 1000 $m^2$ |
|---|---|
| Transmission range | 250m |
| Propagation model | Two-ray ground |
| MAC protocol | 802.11 CSMA (RTS/CTS disabled) |
| Link bandwidth | 2 Mbps |
| Traffic type | CBR (UDP) |
| Packet size | 512 bytes |
| Number of nodes | 15 or 100 |
| Number of radios | $\leq 3$ |
| Hello interval | 1 second |
| Buffer size | 20 packets |

## 5.2 Tuning AntMesh Parameters

We study two configuration parameters of AntMesh routing algorithm. The first is the parameter $p_0$ which is used in our pseudo-random node transition rule and it governs the relative behavior of smart ant forwarding. The parameter $p_0$ determines the probability of choosing a next hop for path discovery/data forwarding with maximum pheromone value or selecting the next hop based upon probability distributions using Eq. (3). If we set the value of $p_0$ to be very high, it would select links with high pheromone values more often than others. Similarly, setting the value too low would spread the data traffic around the links and therefore, would fall back to the classical AntNet algorithm. So there exists a relationship and in order to prove this analytical reasoning, figure 4a demonstrates the performance of AntMesh in terms of average network throughput as a function of $p_0$ under varying network traffic. Notice that the network throughput is almost the same for all the value of $p_0$ when there is little traffic on the network. The $p_0$ value starts affecting the algorithm performance when more packets are beginning to pump in the network increasing the network traffic. At that point, the higher the $p_0$ value, the better the throughput it gives, this is because, most of the time, links of good quality (highest pheromone) are selected for data forwarding. Since the network is still not saturated, the interference estimation rule perfectly captures the interference thereby resulting in increasing the pheromone values of links with less interference. However, notice that when the network traffic is high, the gap between the AntMesh throughput with different values of $p_0$ became small. We believe the reason behind this pattern is due to the fact that since the network is saturated, the low $p_0$ value (i.e. 0.2, 0.5) starts spreading the data over multiple links with more chances of random next hop selection. However, for higher values of $p_0$, AntMesh still gives better throughput than when a lower value is set and therefore, in the rest of our simulations, we will select the value of parameter to $p_0$=0.8.

The second parameter we intend to tune is the smart ant generation rate which is related to AntMesh stability. The stability of any routing algorithm depends upon how quickly it adapts itself to the changing dynamics of the network. We define the time it takes for routing algorithm to learn the best routing policies as the learning time (convergence time). We demonstrate the stability of our algorithm by measuring *path latency* in mesh networks under dynamic load situations when traffic characteristics change or load on the network is increased. In order to effectively demonstrate the quality of the learned policy, we have used the average packet delay as the evaluation metric.

Figure 4b shows the learning times of AntMesh algorithm under different smart ant generation rates. It relates the protocol's stability in the adaptation process for different ant rates during the simulation run time. We have used a 15-node grid mesh topology shown in figure 3 as the underlying network. In the beginning, one flow is generated to introduce a light load on the network. Then at $t = 10s$, 3 more flows are initiated in order to increase the traffic in the network. At $t = 20s$, these 3 flows are stopped in order to bring the network back to its normal state. It can be seen in figure 4b that AntMesh adapts to this increase in network load at $t=10s$, by switching to a new path between the source and destination node with the help of its link and path estimation modules defined in interference estimation rule. However, at $t = 20s$, the algorithm starts converging back to the previous best path that it had discovered. We believe that the convergence of AntMesh to the best available path is related to how fast smart ants are being generated by each node i.e. the ant generation rate. Figure 4b shows that a more frequent ant generation leads to less the time for the algorithm to find a better path. In other words, the time window for finding this path decreases with increasing number of smart ants in the network. However, these values of various ant rate is highly dependent upon the traffic characteristics as well as the network topology and therefore should not be taken as general criteria for applying AntMesh in WMNs.

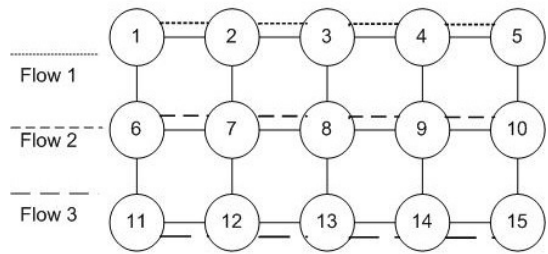

Fig 3. Grid Topology

Although the learning time of AntMesh is inversely proportional to the smart ant generation rate, a higher ant rate means increase control traffic in the network which in turn would limit the overall useful network capacity. Therefore there exists a tradeoff between the amount of routing traffic generated and the learning time. We investigate this tradeoff in AntMesh by using normalized routing load (NRL) as the evaluation metric which is the ratio of the control traffic in the network to the total amount of packets that were actually pumped into the network. Figure 4c shows the NRL when the network is lightly load, and when it is overload under varying smart ants' rates. One interesting thing to note in here is that the network load has almost negligible effect on the routing load when there are too few number of smart ants in the network i.e. the ant generation rate is small. Even for higher ant rates, one can observe that ant generation rates have a marginal effect on the NRL of our routing protocol under the conditions when the network is below saturated. On the other hand, there is a significant difference among the NRL values on the network when the network is overloaded. One reason behind this increase in NRL is due to the increase in packet collisions and the dropping of packets from the node queues when the network becomes overloaded, which in turn will generate more smart ants to discover new paths. Furthermore, there is an extra overhead in communicating queue sizes (2-hop neighbor information) with the neighboring nodes by HSA on each node to compute the inter-flow interference. Although AntMesh has a higher NRL as compared to the overhead of other routing protocols (i.e., distance vector or link state as shown in [12)], we argue that this can be compensated by the efficient performance it provides in routing the traffic by finding less interference paths in WMNs as shown in our results section.

Figure 4d shows the impact on the network performance in terms of end-to-end delay which is depicted as a function of ant generation rates for different network loads. Notice, that the packet delay remains almost the same for increasing number of smart ants in the network.. This is because most of the traffic stays on the best path which keeps the packet delay almost constant among different network load conditions. It is because of this reason, for the rest of our simulations, we select the ant generation rate to 40 ants per second.

### 5.3 Results from WMNs - Stationary Nodes

We want to observe the effectiveness of our proposed smart ant-based routing algorithm in capturing inter and intra-flow interference on a mesh network having multiple radios each configured to multiple channels. For evaluating AntMesh in a multi-radio WMN environment, we modified the node models in our studies to contain one to three radios, where each radio may be configured to work with multiple channels. 802.11b DCF with RTS/CTS disabled is used as the underlying MAC protocol with link bandwidth of 2Mbps. We have run our experiments on an infrastructure grid mesh topology (as in Figure 3) with random location nodes added. In this semi-random topology we place 20 nodes uniform randomly in a 1000m by 1000m area, and four traffic flows are generated destined to 1 to 4 internet gateway nodes.

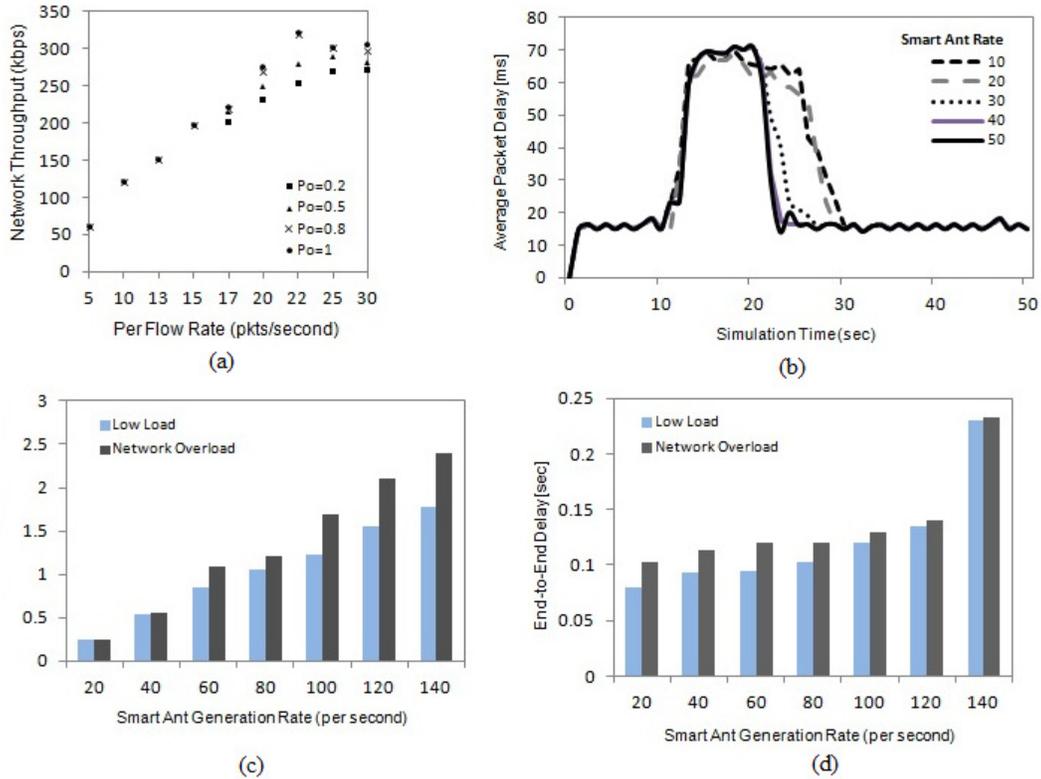

Fig. 4. AntMesh behavior (a) throughput as a function of $p_0$. (b) learning time with various smart ant rates. (c) Effect of smart ant generation on protocol overhead. (d) Effect of smart ant generation on AntMesh end-to-end delay

Figure 5a presents the total network throughput vs. the flow rate for our proposed smart ant-based routing algorithm. We can observe that AntMesh provides better throughput, outperforming the other schemes by as much as 35% especially when the network becomes congested. Both the SARA and DAR perform poor against AntMesh but almost equally with each other since it does not capture the traffic load among the nodes. The increased throughput of AntMesh is due to the capturing of inter- and intra-flow interference by its interference estimation rule. Although, DAR performs close to AntMesh when the network load is light, as the network starts to saturate, nodes' queues start filling up, experiencing more contention on the outgoing links thus affecting the link quality; this is only recognized by smart ants with the help of their path estimation module (PEM) of interference estimation rule (by taking into consideration the queue sizes being periodically sent by HSA to all the neighbors of node). The involvement of the PEM module of AntMesh results in earlier detouring of the flow than with the other schemes. Clearly, the benefit of considering the queue sizes of neighboring nodes to measure the actual interference and traffic load is evident from the increased throughput in the graph.

Similarly, Figure 5b demonstrates the average end-to-end delay. AntMesh outperforms other schemes in terms of end-to-end delay of packets by as much as 30%. Notice though, that the end-to-end delay of DAR approached very close to AntMesh when the network is heavily loaded; this is because of overhead experienced by AntMesh due to the path change captured by the stochastic behaviour of queue lengths.

The packet loss ratios of all three approaches are shown in Figure 5c. When the network load is low, the packet loss ratios, as expected, for all of the three schemes are low; as the network becomes more and more saturated, AntMesh outperforms SARA and DAR by avoiding inter- and intra-flow interference paths. Particularly, the difference in packet loss ratio among AntMesh and other two schemes is very evident and can result in as much as a 60% reduction.

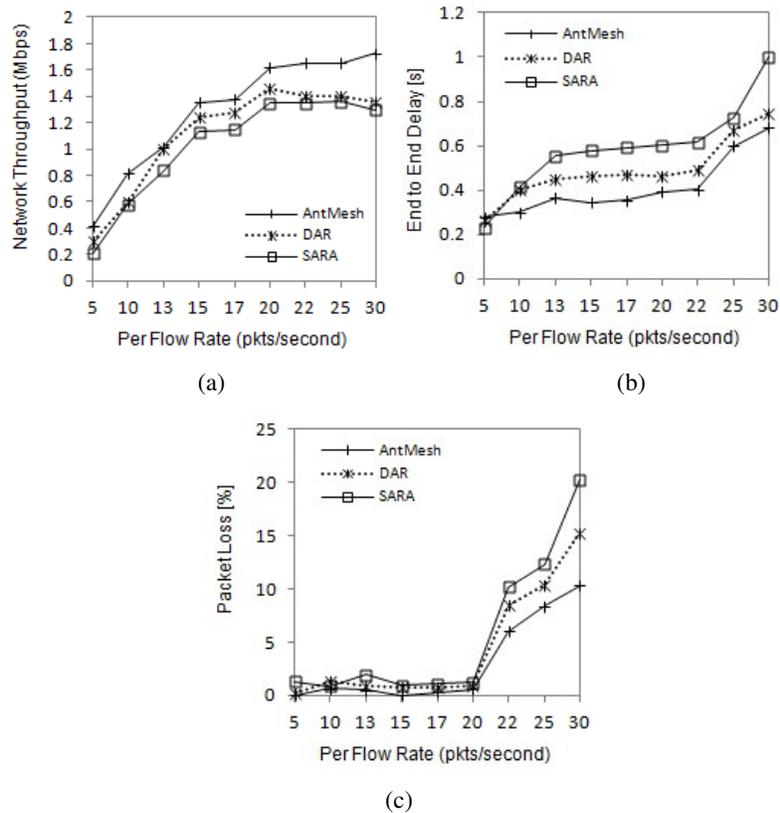

(a)   (b)

(c)

Fig. 5. AntMesh (a) Network throughput. (b) End-to-End delay. (c) Packet Loss

### 5.4 Results from WMNs - Mobile Nodes

In this part of our evaluation, we test the performance of AntMesh in mobile WMNs scenario with static and mobile nodes. We consider a network of 100 nodes randomly distributed in a 500m x 500m region. All the mobile nodes in each simulation run of 60s are configured with the same speed and the random waypoint model [33]. The data traffic consists of 6 CBR flows of random source destination pair. In order to evaluate the effectiveness of our smart ant-based routing algorithm, we have used packet delivery fraction and average end-to-end delay as evaluation metrics. Figures 6a and 6b show the delivery ratio and average end-to-end delay as a function of different node speeds ranging from 0 m/s to 30 m/s. As it can be easily seen from the delivery ratio graph, AntMesh outperforms SARA and DAR clearly and this performance gap between the schemes is more evident when the nodes are moving at a faster rate. However, this performance gap for average packet delay in Figure 6b is less but again it increases for higher nodes' speeds. Notice the dip in the delivery ratio for all the algorithms in Figure 6a, we believe that this is because since nodes are moving in fast speeds, it is possible that some nodes get out of reach from the rest of the network and therefore packets cannot be delivered to them resulting in possible low delivery ratio. Similar arguments can be made for the sharp rise in packet delay of the algorithms in Figure 6b. Furthermore, in the mobility model (RWP) that we have used in our experiments, nodes tend to make sudden and uncorrelated changes in their movement direction at the pause points, which is captured timely by our adaptive smart ant-based routing algorithm resulting in improved performance than the SARA and DAR protocols which are reactive in nature. Another interesting result is that SARA performs better than DAR and very close to AntMesh in both the graphs. This is due to the design objective of SARA which is targeted for networks having highly mobile nodes with critical connectivity. However, it still under performs than AntMesh because of its inability in effectively measuring the interference.

In order to simulate the true nature of a wireless mesh network with mobile nodes, we have collected another set of results by gradually increasing the number of mobile nodes in the network among the fixed nodes in the networks. The results are collected with 20% to 100% of the total nodes in the network given a random movement and direction while the remaining was stationary. We have provided the speed as a simulation parameter that was fixed to 10 m/s. Figures 6c and 6d show the same performance measures for all algorithms as a function of number of mobile nodes in the network. It can be seen from the graphs that AntMesh exhibits a better performance in terms of higher delivery ratio and less packet delay than SARA and DAR particularly when the network is highly mobile. The stochastic forwarding of smart ants for path exploration and timely capture of inter and intra flow interference by our link and path estimation modules of smart ants to pay off more when the number of mobile nodes in the network are increased. The reason for the comparatively poor performance of SARA and DAR lies in the fact that since it uses minimum hop count metric for path construction, therefore, it fails to cope up with the sudden and dynamic changes in the network until the path gets broken which would eventually result in increasing the packet delay due to frequent path reconstructions as shown in Figure 6d. Also, notice that in these graphs too, the performance of SARA is better than that of DAR because of the previously mentioned design objective of SARA.

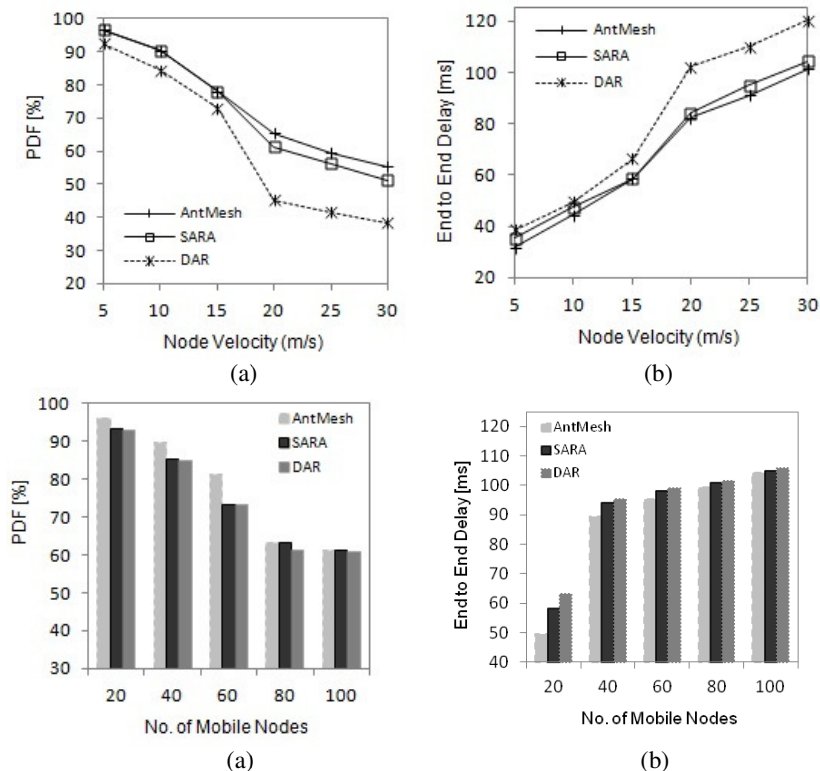

Fig. 6. AntMesh (a) PDF as a function of node speed (b) Packet Delay as a function of node speed (c) PDF as a function of mobile nodes (d) Packet Delay as a function of mobile nodes

## 6. CONCLUSION

This paper studied the problem of packet routing in wireless mesh networks with a specific emphasis on a framework using *smart ants*. To enable the use of such agents we proposed an interference-aware data forwarding scheme called AntMesh which provides a distributed, stochastic heuristic to solve a dynamic network routing problem. In addition, we also emphasized on the importance of having certain desirable properties that smart ants should possess in order to effectively utilize the space/channel diversity typically common in such networks. We demonstrated the stability of AntMesh through simulations that it quickly converges to the best path under situations when traffic characteristics change (among others when load on the network is increased). We have shown that with an appropriate tuning of the parameters, AntMesh behaves better when compared to other competing approaches in mesh networks. In addition, the promising results shown in the paper underline the need for a real-life testbed evaluation on which we are currently working on.

**Authors**

**Fawaz Bokhari**

received his master's degree in computer science from Lahore University of Management Sciences (LUMS) in 2007 and is currently pursuing his Ph.D. studies in computer science from University of Texas at Arlington. His research interests include broadband wireless networking, network and MAC layer designs, wireless multimedia communication, network monitoring and deployment. He is a Fulbright scholar and was awarded this scholarship in 2007 for his PhD studies in United States.

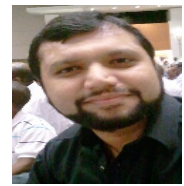

**Gergely Záruba**

received his M.Sc. degree in computer engineering (technical informatics) from the Technical University of Budapest in 1997. He received his Ph.D. in computer science from The University of Texas at Dallas in 2001. He is currently an Associate Professor of computer science and engineering at The University of Texas at Arlington. He is a senior member of the IEEE.

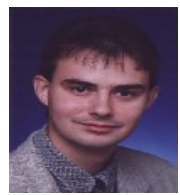